\documentclass[aps,prl,twocolumn,groupedaddress,superscriptaddress]{revtex4}

\usepackage[utf8]{inputenc}
\usepackage[T1]{fontenc}

\usepackage{times}

\usepackage{amssymb,amsfonts,amsmath,amsthm}

\usepackage{slashed}
\usepackage{dcolumn}
\newcolumntype{l}[1]{D{.}{\cdot}{#1}} 

\usepackage{graphicx}
\usepackage[tight,hang,raggedright,normalsize]{subfigure}

\usepackage[
            pdftitle={Nucleon distribution amplitudes from lattice QCD},
            pdfauthor={Nikolaus Warkentin},
            pdfsubject={Calculation of moments of nucleon distribution amplitudes.},
            pdfkeywords={moments, hadron, nucleon, baryon, quark, distribution,wave,function},
            pdfpagemode=UseOutlines,
            colorlinks=true,
            pdfproducer={University of Regensburg},
            ]{hyperref}

\newcommand{\msb}{$\overline{\text{MS}}$}

\begin{document}

\title{Nucleon distribution amplitudes from lattice QCD}

\author{Meinulf \surname{G\"ockeler}}
    \affiliation{Institut f\"ur Theoretische Physik, Universit\"at Regensburg, 93040 Regensburg, Germany}
\author{Roger \surname{Horsley}}
    \affiliation{School of Physics, University of Edinburgh, Edinburgh EH9~3JZ, UK}
\author{Thomas \surname{Kaltenbrunner}}
    \affiliation{Institut f\"ur Theoretische Physik, Universit\"at Regensburg, 93040 Regensburg, Germany}
\author{Yoshifumi \surname{Nakamura}}
    \affiliation{Deutsches Elektronen-Synchrotron DESY and John von Neumann Institut f\"ur
Computing NIC, 15738 Zeuthen, Germany}
\author{Dirk \surname{Pleiter}}
    \affiliation{Deutsches Elektronen-Synchrotron DESY and John von Neumann Institut f\"ur
Computing NIC, 15738 Zeuthen, Germany}
\author{\firstname{Paul} E.~L. \surname{Rakow}}
    \affiliation{Theoretical Physics Division, Department of Mathematical Sciences, University of Liverpool, Liverpool L69~3BX, UK}
\author{Andreas \surname{Sch\"afer}}
    \affiliation{Institut f\"ur Theoretische Physik, Universit\"at Regensburg, 93040 Regensburg, Germany}
    \affiliation{Yukawa Institute for Theoretical Physics, Kyoto University, Japan}
\author{Gerrit \surname{Schierholz}}
    \affiliation{Deutsches Elektronen-Synchrotron DESY and John von Neumann Institut f\"ur
Computing NIC, 15738 Zeuthen, Germany}
\author{Hinnerk~St\"uben}
    \affiliation{Konrad-Zuse-Zentrum f\"ur Informationstechnik Berlin, 14195 Berlin, Germany}
\author{Nikolaus \surname{Warkentin}}
    \affiliation{Institut f\"ur Theoretische Physik, Universit\"at Regensburg, 93040 Regensburg, Germany}
    \email{nikolaus.warkentin@physik.uni-regensburg.de}
\author{\firstname{James} M. \surname{Zanotti}}
    \affiliation{School of Physics, University of Edinburgh, Edinburgh EH9~3JZ, UK}

\collaboration{QCDSF Collaborations} 
\noaffiliation

\date{\today}

\begin{abstract}
We calculate low moments of the leading-twist and next-to-leading twist nucleon
distribution amplitudes on the lattice using two flavors of clover fermions. The
results are presented in the \msb\ scheme at a scale of $2\,\mathrm{GeV}$ and
can be immediately applied in phenomenological studies. We find that the
deviation of the leading-twist nucleon distribution amplitude from its asymptotic
form is less pronounced than sometimes claimed in the literature.

\end{abstract}


\pacs{12.38.Gc, 14.20.Dh}
\preprint{DESY 08-039, Edinburgh 2008/14, LTH 787}

\keywords{nucleon wave function; nucleon distribution amplitude; lattice QCD}

\maketitle

\phantomsection
\addcontentsline{toc}{subsubsection}{Introduction.}
\paragraph{Introduction. ---}
Distribution amplitudes 
\cite{Chernyak:1977as,Efremov:1979qk,Lepage:1979za,Lepage:1980fj,Chernyak:1984bm,Chernyak:1987nu} 
describe  the structure of hadrons in terms of valence quark Fock states at small transverse separation
and are required in the calculation of (semi)exclusive processes.
A simple picture is obtained at very large values of the momentum transfer. 
For example, the magnetic Sachs form factor of the nucleon $G_M(Q^2)$ can then be expressed as a convolution of the hard scattering kernel $h(x_i, y_i, Q^2)$ and the leading-twist quark distribution amplitude in the nucleon $\varphi(x_i, Q^2)$~\cite{Lepage:1979za}:
   \begin{equation*}
\begin{split}
      G_M(Q^2)&
\\
=f_N^2&  \int_0^1 [\mathrm d x] \int_0^1  [\mathrm d y]\;
               \varphi^\star(y_i,Q^2) h(x_i, y_i, Q^2) \varphi(x_i,Q^2)
\end{split}
   \end{equation*}
where $[\mathrm d x]=\mathrm d x_1\mathrm d x_2 \mathrm d x_3\delta(1-\sum_{i=1}^3 x_i)$, and $-Q^2$ is the squared momentum transfer in the hard scattering process. However, in the kinematic region
$1 \, \mathrm{GeV}^2 < Q^2 < 10 \, \mathrm{GeV}^2$, which has attracted a lot
of interest recently due to the JLAB data \cite{Gayou:2001qd,Punjabi:2005wq} for $G_M$, 
the situation is more complicated. Here calculations
are possible, e.g., within the light-cone sum rule approach \cite{Braun:2001tj,Lenz:2003tq}. 
They indicate that higher-twist distribution amplitudes become important while higher Fock 
states do not play a significant role. In any case,
the distribution amplitudes are needed as input.

Being typical nonperturbative quantities, distribution amplitudes are
difficult to compute reliably in a model-indepen\-dent way. Determinations
by QCD sum rules have been attempted, but suffer from considerable
systematic uncertainties, especially for lower values of $Q^2$.
As advocated in the pioneering work \cite{Martinelli:1988xs}, lattice QCD can
provide valuable additional information.

In this paper we present an improved and extended lattice analysis of the nucleon
distribution amplitudes. We find that the asymmetry of the leading-twist amplitude is smaller than in QCD
sum rule calculations, in agreement with phenomenological estimates \cite{Braun:2006hz,Bolz:1996sw},
which suggest a less asymmetric form.

\phantomsection
\addcontentsline{toc}{subsubsection}{General Framework.}
\paragraph{General Framework. ---}

In the case of the proton, the starting point is the matrix element of a trilocal quark operator,
\begin{widetext}
\begin{equation}
      \begin{split}
      \langle 0 \vert &
      \left[
         \exp \left( ig \int_{z_1}^{z_3} A_\mu(\sigma) \mathrm d \sigma^\mu  \right)
         u_\alpha (z_1)
      \right]^a
      \left[
         \exp \left( ig \int_{z_2}^{z_3} A_\nu(\tau) \mathrm d \tau^\nu  \right)
         u_\beta(z_2)
      \right]^b \,
      d_\gamma^c(z_3)\;
      \vert p \rangle \epsilon^{abc}\\
      =&\frac{1}{4} f_N 
      \left\{
            (p\cdot\gamma C)_{\alpha\beta} (\gamma_5 N)_\gamma V(z_i\cdot p)+ (p\cdot \gamma \gamma_5 C)_{\alpha\beta} N_\gamma A(z_i\cdot p)
            +(i\sigma_{\mu\nu}p^\nu C)_{\alpha\beta}(\gamma^\mu\gamma_5 N)_\gamma T(z_i\cdot p)
      \right\}+\text{higher twist},
      \end{split}
      \label{eq_dadef}
   \end{equation} 
\end{widetext}
where path ordering is implied for the exponentials, $a,b,c$ are the color
indices, $N$ the proton spinor and $\vert p \rangle$ denotes a proton state
with momentum $p$. We will consider this matrix element for space time
separation of the quarks on the light cone $z_i=a_i z$ ($z^2=0$) and
$\sum_i a_i=1$.

In momentum space we have
   \begin{equation}
      V(x_i)\equiv\int V(z_i\cdot p) \prod_{i=1}^{3}  \exp\left(ix_i (z_i\cdot p)\right) \frac{\mathrm d(z_i\cdot p)}{2\pi}
      \label{eq_daft}
   \end{equation}
with $V(x_i)\equiv V(x_1,x_2,x_3)$ and similarly for $A(x_i)$ and $T(x_i)$. The distribution amplitudes $V(x_i)$, $A(x_i)$ and $T(x_i)$ describe the quark distribution inside the proton as functions of the longitudinal momentum fractions $x_i$. The dependence on the renormalization scale is suppressed for notational simplicity.

We consider the moments of distribution amplitudes, which are defined as
   \begin{equation}
      V^{lmn}=\int_0^1 [\mathrm dx]\; x_1^l x_2^m x_3^n\; V(x_1, x_2, x_3),
   \end{equation} 
with analogous definitions for the other distribution amplitudes. Using eqs.~(\ref{eq_dadef}) and (\ref{eq_daft}) one can relate the moments of the leading-twist nucleon distribution amplitudes to matrix elements of the local operators

\begin{equation}
    \begin{split}
      & \mathcal V_\tau^{\rho \bar l \bar m \bar n }(0)\equiv 
        \mathcal V_\tau^{\rho (\lambda_1\cdots\lambda_l ) (\mu_1\cdots\mu_m) (\nu_1\cdots\nu_n) }(0) 
                \\
        &               =                     [i^l \mathcal D^{\lambda_1}\dots \mathcal D^{\lambda_l}u(0)]^a_\alpha 
                                                                  (C\gamma^\rho)_{\alpha\beta}
                                                 [i^m \mathcal D^{\mu_1}\dots \mathcal D^{\mu_m} u(0)]^b_\beta \\
        &  \times [i^n \mathcal D^{\nu_1}\dots \mathcal D^{\nu_n} \gamma_5 d(0) ]^c _{\tau} \;\epsilon^{a b c}
    \end{split}
\nonumber
\end{equation} 
by $ \langle 0\vert \mathcal V_\tau^{\rho \bar l \bar m \bar n }(0) \vert p\rangle
=-f_N p^{\rho \bar l \bar m \bar n} N_\tau V^{ l m n}$
with similar relations \cite{Martinelli:1988xs} for the operators $\mathcal
A_\tau^{\rho \bar l \bar m \bar n }$ and $\mathcal T_\tau^{\rho \bar l \bar m
\bar n }$ corresponding to the moments $A^{lmn}$ and $T^{lmn}$, respectively.
Here the multi-indices $\bar l \bar m \bar n$ denote the Lorentz structure
connected with the covariant derivatives on the r.h.s.

Due to the presence of two {\it up}-quarks in the proton and isospin symmetry, the three different amplitudes can be expressed in terms of the single amplitude $\phi(x_i)$ with the corresponding moments
   \begin{equation}
   \phi^{lmn}=\frac{1}{3}(V^{lmn}-A^{lmn}+2T^{lnm}). \label{eq_phidef}
   \end{equation} 
The normalization constant $f_N$ is defined by the choice $\phi^{000}=1$. The moments of the combination $\varphi(x_i)=V(x_i)-A(x_i)$, usually used in sum rule calculations, can easily be obtained as
$\varphi^{l m n}=2\phi^{l m n}-\phi^{ n m l}$.
In the numerical calculation, however, we prefer the combination $\phi^{lmn}$ as the corresponding statistical errors are smaller by a factor of about $3$.
Note that momentum conservation implies
\begin{equation}
 \phi^{lmn}=\phi^{(l+1)mn}+\phi^{l(m+1)n}+\phi^{lm(n+1)}.
 \label{eq_momconserv}
\end{equation}
In particular we have
\begin{equation}
     \begin{split}
          1=&\;\phi^{100}+\phi^{010}+\phi^{001}\\
           =&\;\phi^{200}+\phi^{020}+\phi^{002}+2(\phi^{011}+\phi^{101}+\phi^{110}).
     \end{split}
 \label{eq_srconstraint}
\end{equation}
In the limit of $Q^2\rightarrow \infty$ one gets $\phi(x_i)=120x_1x_2x_3$ \cite{Lepage:1980fj} and the moments $\phi^{l m n}$ are known exactly: $\phi^{100}=\phi^{010}=\phi^{001}=1/3$, $\phi^{200}=\phi^{020}=\phi^{002}=1/7$ and $\phi^{011}=\phi^{101}=\phi^{110}=2/21$.
Thus asymmetries of the type $\phi^{100}-\phi^{010}$ are important
quantities at low energies as they describe the deviation from the asymptotic
case.

In the case of the next-to-leading twist distribution amplitudes we restrict 
ourselves to operators without derivatives, i.e., to the lowest moments. Thus
the problem is simplified greatly since the Lorentz decomposition of the
relevant matrix element involves only two additional constants  $\lambda_1$ and
$\lambda_2$ \cite{Braun:2000kw}. They describe the coupling to the proton of
two independent proton interpolating fields used in QCD sum rules. One of the
operators, $\mathcal L_\tau$, was suggested in \cite{Ioffe:1981kw} and the
other, $\mathcal M_\tau$, in \cite{Chung:1981cc}:
\begin{align}
      \mathcal L_\tau (0)&=
      \epsilon^{a b c}  \left[ {u^a}^T(0)  C\gamma^\rho u^b(0) \right ]\times (\gamma_5 \gamma_\rho d^c(0))_\tau,\nonumber\\
      \mathcal M_\tau (0)&=
      \epsilon^{a b c}  \left[ {u^a}^T(0)  C\sigma^{\mu\nu} u^b(0) \right ]\times (\gamma_5 \sigma_{\mu\nu} d^c(0))_\tau.\nonumber
   \end{align}
Their matrix elements are given by
   \begin{align}
      \langle 0 \vert  \mathcal L_\tau (0) \vert p\rangle &=\lambda_1 m_N N_\tau, \label{eq_lambda1matrix}\\
      \langle 0 \vert  \mathcal M_\tau (0) \vert p\rangle &=\lambda_2 m_N N_\tau.\label{eq_lambda2matrix}
   \end{align}
Due to Fierz identities we have 
\begin{equation}
 (2\lambda_1+\lambda_2)N=\frac{8}{m_N} \langle 0\vert \epsilon^{abc} \left({u^a}^T C d^b\right) \gamma_5 u^c \vert p\rangle,
 \label{eq_lsum}
\end{equation} 
which vanishes in the nonrelativistic limit.

\phantomsection
\addcontentsline{toc}{subsubsection}{Computation.}
\paragraph{Computation. ---}


\begin{table}
\renewcommand{\arraystretch}{1.25}
\footnotesize
     \begin{tabular}{ | c|| D{.}{.}{16} | D{.}{.}{16} |}
\hline
                                             & \multicolumn{1}{c|}{$\beta=5.29$}& \multicolumn{1}{c|}{$\beta=5.40$}    \\ \hline

$ f_N\cdot 10^{3} [\mathrm{GeV}^2]$          & 2.984(60)(157)(65)   & 3.144(61)(29)(54)   \\
$-\lambda_1\cdot 10^{3} [\mathrm{GeV}^2]$    & 39.69(76)(259)(124)  & 38.72(70)(43)(106)  \\
$ \lambda_2\cdot 10^{3} [\mathrm{GeV}^2]$    & 78.70(155)(562)(245) & 76.23(139)(84)(207) \\
\hline
  $\phi^{100}$                               & 0.3549(11)(61)(2)    & 0.3638(11)(68)(3) \\
  $\phi^{010}$                               & 0.3100(10)(73)(1)    & 0.3023(10)(42)(5) \\
  $\phi^{001}$                               & 0.3351(9)(11)(2)     & 0.3339(9)(26)(2)  \\
  $\phi^{100}-\phi^{001}$                    & 0.0199(23)(46)(4)    & 0.0300(23)(93)(1) \\
  $\phi^{001}-\phi^{010}$                    & 0.0251(16)(84)(3)    & 0.0313(17)(12)(7) \\
\hline                                                              
  $\phi^{011}$                               & 0.0863(23)(97)(74)   & 0.0724(18)(82)(70)  \\
  $\phi^{101}$                               & 0.1135(23)(3)(33)    & 0.1136(17)(32)(21)  \\
  $\phi^{110}$                               & 0.0953(21)(58)(31)   & 0.0937(16)(3)(38)   \\
  $\phi^{200}$                               & 0.1508(38)(213)(64)  & 0.1629(28)(7)(68)   \\
  $\phi^{020}$                               & 0.1207(32)(43)(56)   & 0.1289(27)(37)(51)  \\
  $\phi^{002}$                               & 0.1385(36)(47)(64)   & 0.1488(32)(77)(73)  \\
  $\phi^{110}-\phi^{011}$                    & 0.0075(33)(69)(44)   & 0.0211(27)(78)(32)  \\
  $\phi^{101}-\phi^{110}$                    & 0.0172(29)(82)(57)   & 0.0204(21)(134)(50) \\
  $\phi^{200}-\phi^{020}$                    & 0.0335(43)(26)(78)   & 0.0321(33)(69)(55)  \\
  $\phi^{002}-\phi^{020}$                    & 0.0170(36)(8)(56)    & 0.0193(24)(32)(42)  \\
\hline                                                              
\end{tabular}                                                       
\caption{ \label{tab_results}                                       
Moments and asymmetries in the \msb\ scheme at $2\,\mathrm{GeV}$    
 for $\phi^{lmn}=(V^{lmn}-A^{lmn}+2T^{lnm})/3$.
The first error is statistical, the second (third) error represents the uncertainty due to the chiral extrapolation (renormalization).
The systematic errors should be considered with due caution, see the text for their determination.
} 
\end{table}
The required matrix elements between the vacuum and the proton state
are extracted from two-point correlation functions with the investigated
local operators at the sink and a smeared interpolating operator for the
proton at the source. In addition one needs the usual proton correlator
with both source and sink smeared. We have evaluated these two-point 
functions on gauge field configurations generated by the QCDSF/DIK 
collaborations with the standard Wilson gauge action and two flavors 
of nonperturbatively improved Wilson fermions (clover fermions). The 
gauge couplings used are $\beta = 5.29$ and $\beta = 5.40$ corresponding  
to lattice spacings $a \approx 0.075 \, \mbox{fm}$ and 
$a \approx 0.067 \, \mbox{fm}$ via a Sommer parameter of $r_0 = 0.467 \, \mbox{fm}$
\cite{Khan:2006de,Aubin:2004wf}. 
Our smallest pion masses are $380\,\mathrm{MeV}$ ($\beta=5.29$) and 
$420\,\mathrm{MeV}$ ($\beta=5.40$), while the spatial lattice sizes $L$ are 
such that $m_\pi L\geq 3.7$.

Due to the discretization of space-time, the mixing pattern of the
operators on the lattice is more complicated than in the continuum.
It is determined by the transformation behavior of the operators under
the (spinorial) symmetry group of our hypercubic lattice. As operators
belonging to inequivalent irreducible representations cannot mix, we 
derive our operators
from the irreducibly transforming multiplets of
three-quark operators constructed in \cite{Kaltenbrunner:2008pb} in order to reduce
the amount of mixing to a minimum. These irreducible multiplets 
constitute also the basis for the renormalization of our operators,
which is performed nonperturbatively.
To this end we contract our three-quark operators with three quark sources,
amputate the external legs from the resulting four-point functions and impose an
RI$^\prime$-MOM-like renormalization condition. Finally we use continuum
perturbation theory and the renormalization group to convert the results to the
\msb\ scheme at a scale of $2\,\mathrm{GeV}$.  We estimate the corresponding uncertainty by varying the scale at which our renormalization condition is imposed between $10\,\mathrm{GeV}^2$ and $40\,\mathrm{GeV}^2$. In this procedure, the mixing with
``total derivatives'' is automatically taken into account.

In the case of the moments considered in this work we can avoid
the particularly nasty mixing with lower-dimensional operators completely.
Note that the operators $\mathcal V_\tau^{\rho \bar{l} \bar{m} \bar{n}}$,
$\mathcal A_\tau^{\rho \bar{l} \bar{m} \bar{n}}$ and 
$\mathcal T_\tau^{\rho \bar{l} \bar{m} \bar{n}}$ with different 
multi-indices $\rho \bar{l} \bar{m} \bar{n}$ but the same $l m n$ are 
related to the same moments $V^{l m n}$, $A^{l m n}$ and $T^{l m n}$, and
we make use of this fact not only in order to minimize the mixing problems
but also in order to reduce the statistical noise by considering
suitable linear combinations.

\begin{figure}
     \includegraphics[width=0.97\columnwidth,clip]{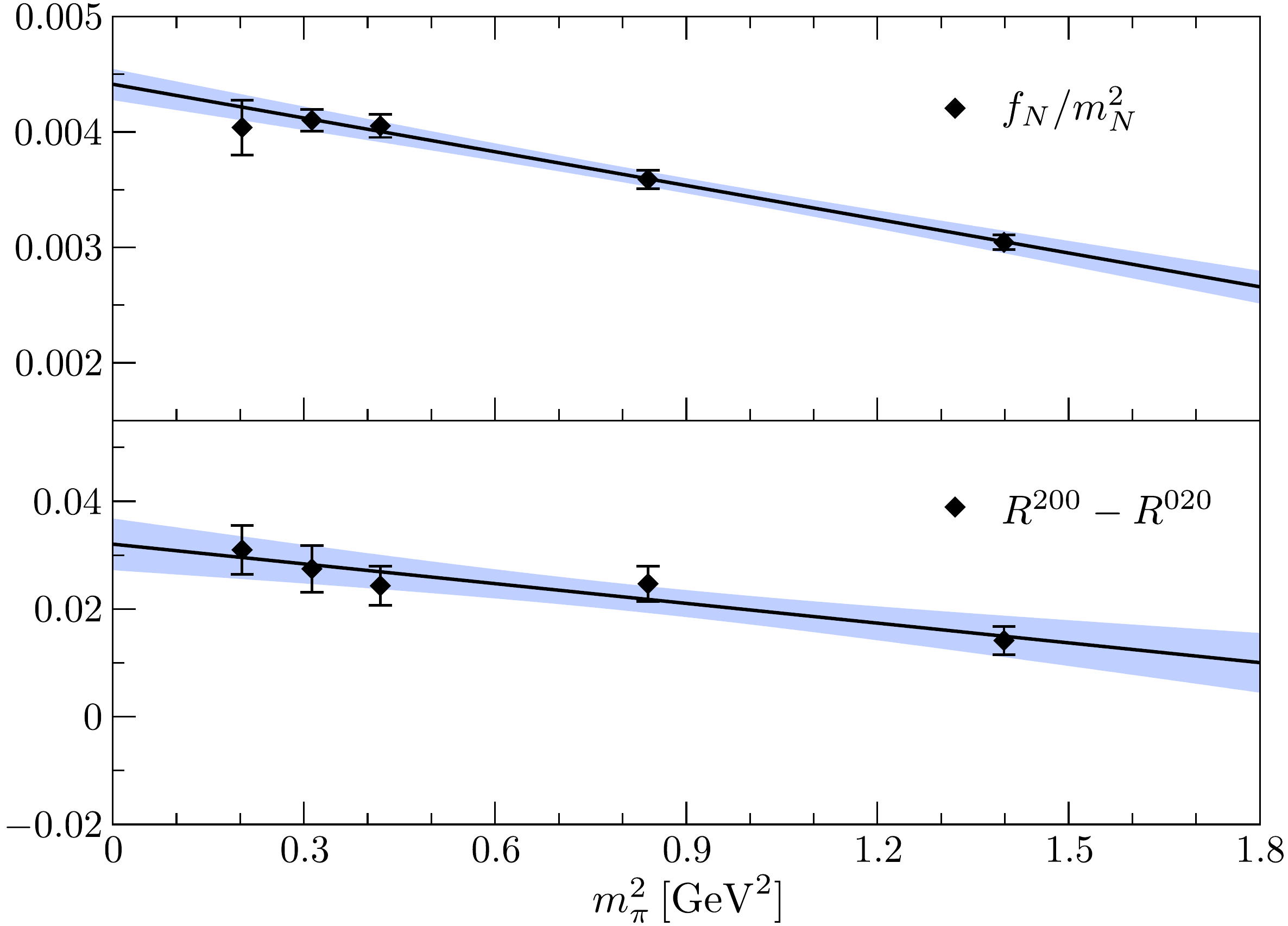}
\caption{\label{fig_mom} Linear chiral extrapolation of bare lattice results for $f_N/m_N^2$ (upper panel) and the asymmetry $R^{200}-R^{020}$ (lower panel) with one-$\sigma$ error band. }
\end{figure}

For the operators without 
derivatives, i.e., the matrix elements $\lambda_1$, $\lambda_2$ and 
$f_N$, we have performed a joint fit of all contributing correlators 
to obtain the values at the simulated quark masses. As these are 
larger than the physical masses a chiral extrapolation to the physical 
point is required in the end. To the best of our knowledge there are no
results from chiral perturbation theory to guide this extrapolation. 
Therefore we have adopted a more phenomenological approach aiming at
linear (in $m_\pi^2$) fits to our data.
It turns out that the ratios $f_N / m_N^2$ and $\lambda_i / m_N$ 
are particularly well suited for this purpose 
(see Fig.~\ref{fig_mom} (upper panel)  for an  example). 
Moreover, $f_N / m_N^2$ is dimensionless and hence not 
affected by any uncertainty in the determination of the lattice spacing.
In order to estimate the systematic error due to our linear extrapolation, we 
also consider a chiral extrapolation including a term quadratic in $m_\pi^2$ and 
take the difference as the systematic error.
The results in the $\overline{\mathrm{MS}}$ scheme at a scale of $2 \, \mathrm{GeV}$ 
are given in Table~\ref{tab_results}. 
Note that $2 \lambda_1 \approx - \lambda_2$, a
relation that is expected to hold in the nonrelativistic limit 
(cf.,\ eq.~\eqref{eq_lsum}).

For the higher moments one can proceed in the same way and the constraint
\eqref{eq_srconstraint} is satisfied very well. However, the
statistical errors in this approach are too large to allow an accurate determination
of the (particularly interesting) asymmetries.  We achieved smaller errors
by calculating the ratios $R^{l m n}= \phi^{l m n}/S_i$ where 
$
        S_1 = \phi^{100}+\phi^{010}+\phi^{001} 
$
 \ for $l+m+n=1$, and
$
 S_2 = 2(\phi^{011}+\phi^{101}+\phi^{110})+\phi^{200}+\phi^{020}+\phi^{002}
$
\ for $l+m+n=2$.
These ratios can be extrapolated linearly to the 
physical masses. An example is shown in Fig.~\ref{fig_mom} (lower panel)
for the case of the 
asymmetry $R^{200}-R^{020}$. Requiring that the constraint \eqref{eq_srconstraint} be satisfied for the 
renormalized values we can finally extract the moments from 
the ratios.

\phantomsection
\addcontentsline{toc}{subsubsection}{Discussion and Conclusions.}
\paragraph{Discussion and Conclusions. ---}

\begin{table}
  \renewcommand{\arraystretch}{1.25}
   \begin{tabular*}{0.95\columnwidth}{ @{\extracolsep{\fill}}| ccc  ||  ccc | ccc |  ccc | ccc|}
\hline
  &            &                           & & LAT   & & &QCDSR& & & BLW & & & BK  &\\
\hline
  &$V^{001}$   &                           & & 0.304 & & &0.248& & &0.303& & &0.311&\\
  &$2 A^{010}$ &                           & & 0.091 & & &0.303& & &0.116& & &0.064&\\
\hline
\end{tabular*}
\caption{\label{tab:comp}
Comparison of our results (LAT) to selected sum rule results \cite{King:1986wi} (QCDSR) and the phenomenological
estimates \cite{Braun:2006hz} (BLW) and \cite{Bolz:1996sw} (BK) at the scale $2\,\mathrm{GeV}$.
} 
\end{table}
Since we only have results at two different lattice spacings, we are unable to 
extrapolate our results to the continuum limit. However, we find that the 
results obtained at $\beta=5.29$ and $\beta=5.40$ are compatible within errors. 
Hence we take the data from our finer lattice ($\beta=5.40$) as our
final numbers. The values for $\phi^{lmn}$ imply that $\varphi^{100}=0.394$,
$\varphi^{010}=0.302$ and $\varphi^{001}=0.304$. These moments
can be interpreted as the fraction of momentum carried by the
corresponding quarks \cite{Chernyak:1984bm,Chernyak:1987nu}.
Thus we find that
the largest fraction of the proton longitudinal momentum is carried by
one {\it up}-quark with spin aligned with the proton spin. However,
this asymmetry is not as strong as found in the QCD sum rule calculation.
Our results for the first moments are close to phenomenological
estimates \cite{Braun:2006hz,Bolz:1996sw}, cf.~Table~\ref{tab:comp}. On the other
hand, our results for $\varphi^{011}$, $\varphi^{101}$ and
$\varphi^{110}$ are similar to the sum rule values, while the
asymmetries in the moments $\varphi^{200}$, $\varphi^{020}$ and
$\varphi^{002}$ are clearly smaller.

\begin{figure}[t]
    \includegraphics[width=0.97\columnwidth,clip]{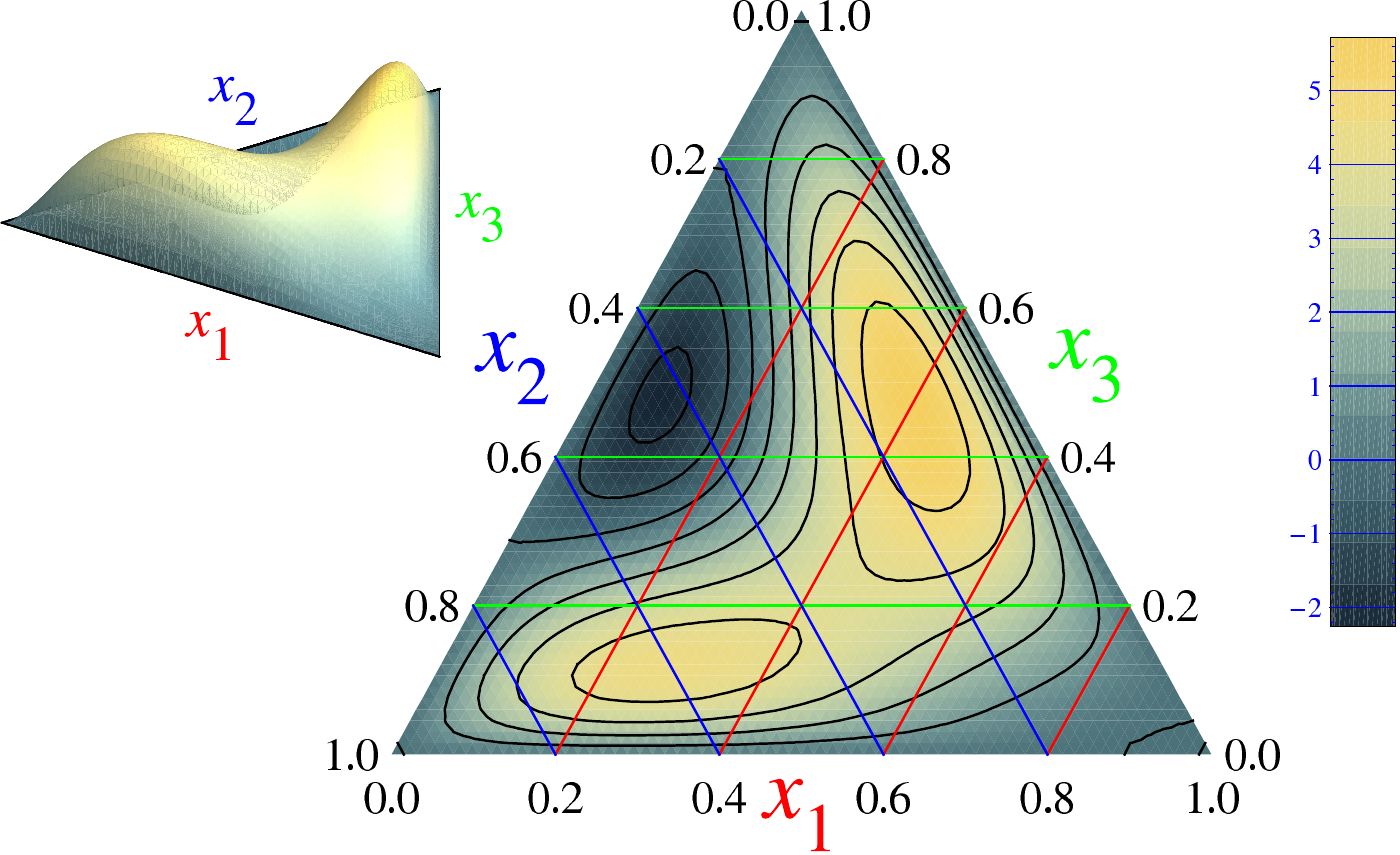}
    \caption{ \label{fig_nda}
    Barycentric contour plot of the leading-twist distribution amplitude
    $\varphi(x_1,x_2,x_3)$ at $\mu=\mu_0=2\,\mathrm{GeV}$
    as obtained from the moments presented in Table~\ref{tab_results}.
    The lines of constant $x_1$, $x_2$ and $x_3$ are parallel 
    to the sides of the triangle labelled by $x_2$, $x_3$ and $x_1$, respectively.
    }
    \end{figure}

\begin{figure}[t]
\includegraphics[width=0.97\columnwidth,clip]{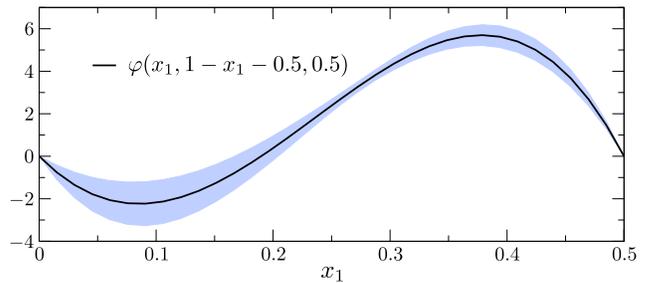}
    \caption{ \label{fig_errorprofile}
    The model distribution amplitude $\varphi(x_1,x_2,x_3)$ for $x_3=0.5$ as a function of $x_1$ with
    statistical errors.
    }
    \end{figure}

Let us now expand the distribution amplitude in terms of polynomials
$P_n$ to order $N=2$ chosen such that the mixing matrix is diagonal
\cite{Braun:1999te,Stefanis:1999wy}:
\begin{equation*}
 \varphi(x_i,\mu)=120 x_1 x_2 x_3 \sum_{n=0}^N c_n(\mu_0) P_n(x_i)
\left(\frac{\alpha_s(\mu)}{\alpha_s(\mu_0)}\right)^{\omega_n}.
\end{equation*}
Calculating the coefficients $c_n(\mu_0)$ from an independent subset of
the moments $\phi^{lmn}(\mu_0=2\,\mathrm{GeV})$, we
obtain a model function for the distribution amplitude presented in
Fig.~\ref{fig_nda}.  
While the (totally symmetric) asymptotic amplitude $120 \, x_1 x_2 x_3$ has a maximum
for $x_1 = x_2 = x_3 = 1/3$, inclusion of the first moments (i.e., choosing $N=1$)
moves this maximum to $x_1 \approx 0.46$,  $x_2 \approx 0.27$,  $x_3 \approx 0.27$
giving the first quark substantially more momentum than the others.
The second moments then turn this single maximum into the two local maxima 
in Fig.~\ref{fig_nda}.
The approximate symmetry in
$x_2$ and $x_3$ is due to the approximate symmetry 
$\varphi^{lmn}\approx \varphi^{lnm}$ of our results. It is also seen 
in QCD sum rule calculations as well as in several 
models such as BLW and BK and may indicate the formation of a diquark system.
To illustrate the statistical uncertainty we show in
Fig.~\ref{fig_errorprofile} the profile of $\varphi$ at $x_3=0.5$ and
the corresponding error band.
Note that higher-order polynomials have been disregarded in this model and thus
Figs.~\ref{fig_nda} and \ref{fig_errorprofile} should be interpreted with
due caution.

In order to establish a link to observable quantities we are calculating nucleon 
form factors using light cone sum rules. As our moments are close to the phenomenological
estimates shown in Table~\ref{tab:comp} we can expect reasonable agreement with the 
experimental data. It is, however, to be stressed that these calculations still involve
some model dependence, while the moments presented in this work were obtained from first
principles.

\paragraph*{Acknowledgment. ---}
\begin{acknowledgments}
We are grateful to A.~Lenz, J.~Bloch, A.~Manashov and V.~Braun for helpful
discussions. The numerical calculations have been performed on the Hitachi
SR8000 at LRZ (Munich), apeNEXT and APEmille at NIC/DESY (Zeuthen) and
BlueGene/Ls at NIC/JSC (J\"ulich), EPCC (Edinburgh) and KEK (by the Kanazawa
group as part of the DIK research program) as well as  QCDOC (Regensburg)
 using the Chroma software library \cite{Edwards:2004sx, bagel:2005}. 
This work was supported by DFG (Forschergruppe Gitter-Hadronen-Ph\"anomenologie), 
by EU I3HP (contract No. RII3-CT-2004-506078) 
 and by BMBF.
\end{acknowledgments}

\providecommand{\href}[2]{#2}\begingroup\raggedright\endgroup


\begin{thebibliography}{10}

\bibitem{Chernyak:1977as}
V.~L. Chernyak and A.~R. Zhitnitsky, ``{Asymptotic Behavior of Hadron
  Form-Factors in Quark Model. (In Russian)},''
{\em JETP Lett.} {\bf 25} (1977)  510.

\bibitem{Efremov:1979qk}
A.~V. Efremov and A.~V. Radyushkin, ``{Factorization and Asymptotical Behavior
  of Pion Form- Factor in QCD},''
\href{http://dx.doi.org/10.1016/0370-2693(80)90869-2}{{\em Phys. Lett.} {\bf
  B94} (1980)  245--250}.

\bibitem{Lepage:1979za}
G.~P. Lepage and S.~J. Brodsky, ``{Exclusive Processes in Quantum
  Chromodynamics: The Form- Factors of Baryons at Large Momentum Transfer},''
\href{http://dx.doi.org/10.1103/PhysRevLett.43.545}{{\em Phys. Rev. Lett.} {\bf
  43} (1979)  545--549}.

\bibitem{Lepage:1980fj}
G.~P. Lepage and S.~J. Brodsky, ``{Exclusive Processes in Perturbative Quantum
  Chromodynamics},''
\href{http://dx.doi.org/10.1103/PhysRevD.22.2157}{{\em Phys. Rev.} {\bf D22}
  (1980)  2157}.

\bibitem{Chernyak:1984bm}
V.~L. Chernyak and I.~R. Zhitnitsky, ``{Nucleon Wave Function and Nucleon
  Form-Factors in QCD},''
\href{http://dx.doi.org/10.1016/0550-3213(84)90114-7}{{\em Nucl. Phys.} {\bf
  B246} (1984)  52--74}.

\bibitem{Chernyak:1987nu}
V.~L. Chernyak, A.~A. Ogloblin, and I.~R. Zhitnitsky, ``{The wave functions of
  the octet baryons},''
\href{http://dx.doi.org/10.1007/BF01557663}{{\em Z. Phys.} {\bf C42} (1989)
  569}.

\bibitem{Gayou:2001qd}
{\bf Jefferson Lab Hall A} Collaboration, O.~Gayou {\em et al.}, ``{Measurement
  of {$G_{Ep}/G_{Mp}$} in {$\vec e p\rightarrow e\vec p$} to
  {$Q^2=5.6\,\textrm{GeV}^2$}},''
  \href{http://dx.doi.org/10.1103/PhysRevLett.88.092301}{{\em Phys. Rev. Lett.}
  {\bf 88} (2002)  092301},
\href{http://arxiv.org/abs/nucl-ex/0111010}{{\tt arXiv:nucl-ex/0111010}}.

\bibitem{Punjabi:2005wq}
V.~Punjabi {\em et al.}, ``{Proton elastic form factor ratios to {$Q^2 =
  3.5~\textrm{GeV}^2$} by polarization transfer},''
  \href{http://dx.doi.org/10.1103/PhysRevC.71.055202}{{\em Phys. Rev.} {\bf
  C71} (2005)  055202},
\href{http://arxiv.org/abs/nucl-ex/0501018}{{\tt arXiv:nucl-ex/0501018}}.

\bibitem{Braun:2001tj}
V.~M. Braun, A.~Lenz, N.~Mahnke, and E.~Stein, ``{Light-cone sum rules for the
  nucleon form factors},''
  \href{http://dx.doi.org/10.1103/PhysRevD.65.074011}{{\em Phys. Rev.} {\bf
  D65} (2002)  074011},
\href{http://arxiv.org/abs/hep-ph/0112085}{{\tt arXiv:hep-ph/0112085}}.

\bibitem{Lenz:2003tq}
A.~Lenz, M.~Wittmann, and E.~Stein, ``{Improved light-cone sum rules for the
  electromagnetic form factors of the nucleon},''
  \href{http://dx.doi.org/10.1016/j.physletb.2003.12.009}{{\em Phys. Lett.}
  {\bf B581} (2004)  199--206},
\href{http://arxiv.org/abs/hep-ph/0311082}{{\tt arXiv:hep-ph/0311082}}.

\bibitem{Martinelli:1988xs}
G.~Martinelli and C.~T. Sachrajda, ``{The quark distribution amplitude of the
  proton: a lattice computation of the lowest two moments},''
\href{http://dx.doi.org/10.1016/0370-2693(89)90874-5}{{\em Phys. Lett.} {\bf
  B217} (1989)  319}.

\bibitem{Braun:2006hz}
V.~M. Braun, A.~Lenz, and M.~Wittmann, ``{Nucleon form factors in QCD},''
  \href{http://dx.doi.org/10.1103/PhysRevD.73.094019}{{\em Phys. Rev.} {\bf
  D73} (2006)  094019},
\href{http://arxiv.org/abs/hep-ph/0604050}{{\tt arXiv:hep-ph/0604050}}.

\bibitem{Bolz:1996sw}
J.~Bolz and P.~Kroll, ``{Modelling the nucleon wave function from soft and hard
  processes},'' \href{http://dx.doi.org/10.1007/s002180050186}{{\em Z. Phys.}
  {\bf A356} (1996)  327},
\href{http://arxiv.org/abs/hep-ph/9603289}{{\tt arXiv:hep-ph/9603289}}.

\bibitem{Braun:2000kw}
V.~Braun, R.~J. Fries, N.~Mahnke, and E.~Stein, ``{Higher twist distribution
  amplitudes of the nucleon in QCD},''
  \href{http://dx.doi.org/10.1016/S0550-3213(00)00516-2}{{\em Nucl. Phys.} {\bf
  B589} (2000)  381--409},
\href{http://arxiv.org/abs/hep-ph/0007279}{{\tt arXiv:hep-ph/0007279}}.

\bibitem{Ioffe:1981kw}
B.~L. Ioffe, ``{Calculation of Baryon Masses in Quantum Chromodynamics},''
\href{http://dx.doi.org/10.1016/0550-3213(81)90259-5}{{\em Nucl. Phys.} {\bf
  B188} (1981)  317--341}.

\bibitem{Chung:1981cc}
Y.~Chung, H.~G. Dosch, M.~Kremer, and D.~Schall, ``{Baryon Sum Rules and Chiral
  Symmetry Breaking},''
\href{http://dx.doi.org/10.1016/0550-3213(82)90154-7}{{\em Nucl. Phys.} {\bf
  B197} (1982)  55}.

\bibitem{Khan:2006de}
A.~{Ali Khan} {\em et al.}, ``{Axial coupling constant of the nucleon for two
  flavours of dynamical quarks in finite and infinite volume},''
  \href{http://dx.doi.org/10.1103/PhysRevD.74.094508}{{\em Phys. Rev.} {\bf
  D74} (2006)  094508},
\href{http://arxiv.org/abs/hep-lat/0603028}{{\tt arXiv:hep-lat/0603028}}.

\bibitem{Aubin:2004wf}
C.~Aubin {\em et al.}, ``{Light hadrons with improved staggered quarks:
  Approaching the continuum limit},''
  \href{http://dx.doi.org/10.1103/PhysRevD.70.094505}{{\em Phys. Rev.} {\bf
  D70} (2004)  094505},
\href{http://arxiv.org/abs/hep-lat/0402030}{{\tt arXiv:hep-lat/0402030}}.

\bibitem{Kaltenbrunner:2008pb}
T.~Kaltenbrunner, M.~G{\"o}ckeler, and A.~Sch{\"a}fer, ``{Irreducible
  Multiplets of Three-Quark Operators on the Lattice: Controlling Mixing under
  Renormalization},''
  \href{http://dx.doi.org/10.1140/epjc/s10052-008-0596-4}{{\em Eur. Phys. J.}
  {\bf C55} (2008)  387--401},
\href{http://arxiv.org/abs/0801.3932}{{\tt arXiv:0801.3932 [hep-lat]}}.

\bibitem{King:1986wi}
I.~D. King and C.~T. Sachrajda, ``{Nucleon Wave Functions and QCD Sum Rules},''
\href{http://dx.doi.org/10.1016/0550-3213(87)90019-8}{{\em Nucl. Phys.} {\bf
  B279} (1987)  785}.

\bibitem{Braun:1999te}
V.~M. Braun, S.~E. Derkachov, G.~P. Korchemsky, and A.~N. Manashov, ``{Baryon
  distribution amplitudes in {QCD}},''
  \href{http://dx.doi.org/10.1016/S0550-3213(99)00265-5}{{\em Nucl. Phys.} {\bf
  B553} (1999)  355--426},
\href{http://arxiv.org/abs/hep-ph/9902375}{{\tt arXiv:hep-ph/9902375}}.

\bibitem{Stefanis:1999wy}
N.~G. Stefanis, ``{The physics of exclusive reactions in QCD: Theory and
  phenomenology},'' {\em Eur. Phys. J. direct} {\bf C7} (1999)  1,
\href{http://arxiv.org/abs/hep-ph/9911375}{{\tt arXiv:hep-ph/9911375}}.

\bibitem{Edwards:2004sx}
{\bf SciDAC} Collaboration, R.~G. Edwards and B.~Jo{\'o}, ``{The Chroma
  software system for lattice QCD},''
  \href{http://dx.doi.org/10.1016/j.nuclphysbps.2004.11.254}{{\em Nucl. Phys.
  Proc. Suppl.} {\bf 140} (2005)  832},
\href{http://arxiv.org/abs/hep-lat/0409003}{{\tt arXiv:hep-lat/0409003}}.

\bibitem{bagel:2005}
P.~A. Boyle, 2005.
\newblock \url{http://www.ph.ed.ac.uk/~paboyle/bagel/Bagel.html}.

\end{thebibliography}

\end{document}